\documentclass[aps,prb,twocolumn,longbibliography,superscriptaddress,article]{revtex4-1}
\usepackage{epsfig}
\usepackage{epstopdf}
\usepackage{amsmath}
\usepackage{amsfonts}
\usepackage{amssymb}
\usepackage{hyperref}
\usepackage{bm}
\usepackage{makecell}
\usepackage{rotating}
\usepackage{hyperref}
\usepackage{multirow}
\usepackage{graphicx}
\usepackage{booktabs}
\usepackage{siunitx}
\usepackage{makecell}
\usepackage{color, soul}
\usepackage{ulem}

\usepackage{graphicx}
\usepackage{dcolumn}
\usepackage{bm}
\usepackage{color}

\usepackage{tikz,xcolor,hyperref}

\definecolor{lime}{HTML}{A6CE39}
\DeclareRobustCommand{\orcidicon}{%
	\begin{tikzpicture}
	\draw[lime, fill=lime] (0,0)
	circle [radius=0.16]
	node[white] {{\fontfamily{qag}\selectfont \tiny ID}};
	\draw[white, fill=white] (-0.0625,0.095)
	circle [radius=0.007];
	\end{tikzpicture}
	\hspace{-2mm}
}

\foreach \x in {A, ..., Z}{%
	\expandafter\xdef\csname orcid\x\endcsname{\noexpand\href{https://orcid.org/\csname orcidauthor\x\endcsname}{\noexpand\orcidicon}}
}

\begin{document}

\title{Berry-Curvature Activation by Loop-Current Order in a Kagome Altermagnet
}

\author{Meysam Bagheri Tagani\orcidA}
\email{mtagani@magtop.ifpan.edu.pl}
\affiliation{International Research Centre Magtop, Institute of Physics, Polish Academy of Sciences, Aleja Lotnik\'ow 32/46, PL-02668 Warsaw, Poland}
\affiliation{Department of Physics, University of Guilan, P. O. Box 41335-1914, Rasht, Iran}

\author{Carmine Autieri\orcidB}
\email{autieri@magtop.ifpan.edu.pl}
\affiliation{International Research Centre Magtop, Institute of Physics, Polish Academy of Sciences,
Aleja Lotnik\'ow 32/46, PL-02668 Warsaw, Poland}

\date{\today}
\begin{abstract}
We investigate Berry-curvature activation and anomalous Hall transport in a
kagome altermagnet with a compensated coplanar $120^\circ$ magnetic texture.
Using a minimal tight-binding model containing nearest-neighbor hopping,
noncollinear exchange coupling, intrinsic spin--orbit coupling, and a
time-reversal-odd loop-current order, we disentangle the magnetic, orbital,
and relativistic mechanisms governing the electronic response. The exchange
field produces pronounced momentum-dependent spin splitting and spin-polarized
Fermi surfaces without generating a net magnetization. Nevertheless, in the
absence of loop-current order, a hidden antiunitary symmetry
$\mathcal{T}C_{2z}$ enforces vanishing Berry curvature and intrinsic anomalous
Hall conductivity, even for finite spin--orbit coupling. A directed imaginary
bond order breaks this protection and activates finite Berry curvature and a
sizable, strongly filling-dependent Hall response already in the
nonrelativistic limit. Spin--orbit coupling subsequently reconstructs the
avoided crossings and redistributes the Berry curvature, enhancing or
suppressing the Hall response depending on filling. For sufficiently strong
loop-current order and spin--orbit coupling, a global gap opens at
$n_e=3$, and the Hall conductivity approaches $2e^2/h$, consistent with an
occupied-band Chern number of magnitude two. Parameter-space and
filling-dependent calculations further demonstrate that the Hall-active
regime extends over broad ranges of exchange coupling, spin--orbit coupling,
and chemical potential and remains robust against symmetry-preserving
longer-range hopping. These results identify orbital-current order as an
independent route for converting a Hall-silent kagome altermagnet into an
anomalous Hall metal or a gapped topological phase without net magnetization,
noncoplanar spin order, or scalar spin chirality.
\end{abstract}


\maketitle
	
\section{Introduction}

The recent identification of altermagnetic order has revealed a fundamentally
new class of magnetic quantum matter in which a vanishing net magnetization
coexists with momentum‑dependent non-relativistic spin-splitting, typically associated with
ferromagnets~\cite{vsmejkal2022emerging, song2025altermagnets, vsmejkal2022beyond, krempasky2024altermagnetic, bai2024altermagnetism, reimers2024direct, ding2024large, yang2025three, Tenzin2025, g32j-hnvz, hirakida2025multipoleanalysisspincurrents, leon2025strainenhancedaltermagnetismca3ru2o7,ssxp-gz9l, bagheri2026quantum, tagani2026ferroelectric}.
In altermagnets, the main properties such as breaking of time-reversal symmetry and anomalous Hall effect\cite{Benny26} come from the collinear part with zero net magnetization, even if spin cantings are allowed in altermagnets in the relativistic limit due to antisymmetric exchange\cite{839n-rckn,PhysRevB.111.054442,Fakhredine26}. 
While non-collinear magnets were excluded from the first definitions of altermagnets, recently, some authors extended the definition of altermagnet to non-collinear materials that break time-reversal symmetry and exhibit zero magnetization in the non-relativistic limit protected by crystal symmetries.\cite{Cheong2024,10.1063/5.0283630,schrade2026altermagneticsuperconductingdiodeeffect,PhysRevB.101.220403}
These noncollinear magnets offer a qualitatively rich route to spin symmetry
breaking and electronic structure engineering~\cite{nakatsuji2015large, lee2025spin, takagi2025spontaneous, rimmler2025non, deng2024phase, yoon2023handedness, takeuchi2021chiral, kimata2019magnetic, nan2020controlling, reichlova2019imaging, liu2018electrical}. In this regard, kagome lattices
occupy a central position: geometric frustration, flat‑band physics, Dirac
singularities, and strong quantum interference naturally enhance
Berry‑curvature effects and quantum‑geometric responses~\cite{ye2018massive, kang2020dirac, ortiz2020cs, li2024spin, li2023field, yin2022topological, di2023flat, cheng2026interwoven}. These features have already placed kagome materials at the forefront of modern condensed‑matter
research, particularly in connection with anomalous Hall transport, Chern
magnetism, unconventional superconductivity, correlated topological
phases~\cite{ortiz2020cs, li2024spin, li2023field, yin2022topological, yin2022topological, di2026kagome, tian2023two} and giant tunneling magnetoresistance\cite{Gurung2024-gn}. We will show that noncollinear kagome altermagnets provide a unified platform in which momentum‑dependent spin
splitting, frustrated magnetism, and topological band geometry emerge
intrinsically from lattice symmetry and magnetic texture alone.

A central theme emerging across modern quantum materials research is the
realization that noncollinear magnetic textures can generate electronic
responses far beyond those permitted in conventional collinear magnets.  In
systems where spin, lattice, and orbital degrees of freedom intertwine, even coplanar magnetic configurations can host symmetry‑protected band degeneracies,
momentum‑space spin–momentum locking, and unconventional Berry curvature
patterns.  Recent studies on frustrated antiferromagnets, chiral spin liquids,
and kagome metals have demonstrated that the combination of geometric
frustration and multi‑sublattice magnetism can stabilize electronic states with
giant anomalous Hall effects, momentum‑selective spin polarization, and
topological band inversions despite the absence of net magnetization~\cite{vsmejkal2018topological, mong2010antiferromagnetic, pramanik2019magnetic, arsenijevic2016anomalous, broholm2020quantum, wietek2015nature}.  These
developments have established noncollinear magnets as a fertile platform for
engineering Berry curvature, exploring hidden antiunitary symmetries, and
realizing topological responses that are inaccessible in both ferromagnets and
collinear antiferromagnets.

Despite this rapid progress, the microscopic mechanisms by which noncollinear kagome magnets generate momentum‑dependent spin splitting and Berry curvature
responses remain far from fully understood.  Most theoretical studies have
focused either on collinear altermagnets, where symmetry analysis is relatively
straightforward, or on strongly spin–orbit‑coupled kagome metals, where the
dominant effects arise from heavy‑element chemistry rather than from the
magnetic texture itself.  By contrast, the regime in which a purely coplanar
$120^\circ$ magnetic order, weak spin–orbit coupling, and lattice‑imposed sublattice asymmetry cooperate to produce a spin‑selective band geometry that has received little attention~\cite{cheong2025altermagnetism}.  In particular, it remains an open question to what extent noncollinear kagome altermagnets can host intrinsic momentum-space spin polarization or Berry-curvature activation when net magnetization is absent and relativistic effects are weak. This
gap highlights the need for a minimal, analytically transparent model that
isolates the role of magnetic texture and lattice symmetry in shaping the
electronic topology of kagome altermagnets.

The importance of addressing this question is underscored by a growing body of
experimental work on kagome metals and frustrated antiferromagnets, where
noncollinear magnetic order, weak spin–orbit coupling, and anomalous Hall
signals coexist in unexpected ways~\cite{di2026kagome, wang2008spin}.  Recent measurements on Mn‑based kagome
compounds, itinerant $120^\circ$ antiferromagnets, and orbital‑flux–driven
kagome systems have revealed large Berry‑curvature responses that cannot be
explained solely by conventional spin chirality or relativistic mechanisms~\cite{samanta2024emergence, garcia2026site}.
These observations suggest that purely orbital or symmetry‑driven mechanisms
may play a decisive role in generating topological transport in frustrated
magnets.  Yet a unified theoretical framework capable of disentangling the
effects of noncollinearity, sublattice structure, and emergent orbital flux is
still lacking.  Developing such a framework is essential for interpreting
ongoing experiments and for guiding the search for topological altermagnetic
materials beyond the paradigm of strong spin–orbit coupling.

In this work, we develop a minimal tight‑binding theory for itinerant electrons
on a kagome lattice hosting a compensated coplanar $120^\circ$ altermagnetic
texture. The model incorporates nearest‑neighbor hopping, noncollinear
exchange coupling, intrinsic spin–orbit interaction, and an emergent chiral
orbital flux that mimics interaction‑driven circulating currents on the kagome
network.  Using symmetry analysis together with momentum‑space Berry curvature
calculations, we demonstrate that the system realizes multiple distinct
electronic regimes.  In the absence of chiral flux, the coplanar magnetic phase
preserves a hidden antiunitary symmetry $\mathcal{T}C_{2z}$, which protects a
Berry‑curvature‑compensated altermagnetic metal despite strong
momentum‑dependent spin splitting.  Remarkably, we show that activating the
orbital chiral flux immediately breaks this compensation mechanism and
generates finite Berry curvature and anomalous Hall conductivity even without
relativistic spin–orbit coupling or scalar spin chirality.  The resulting state
constitutes a nonrelativistic topological altermagnetic metal in which topology
emerges purely from orbital gauge structure and frustrated magnetic symmetry.
Our results establish noncollinear kagome altermagnets as a versatile platform
for engineering spin‑split topological electronic states beyond the
conventional paradigm of ferromagnetism and strong spin–orbit coupling.

\section{Model, Formalism, and Symmetry Analysis}
\label{sec:model}

We investigate itinerant electrons on a kagome lattice hosting a compensated noncollinear altermagnetic texture.  
The minimal tight-binding Hamiltonian is written as
\begin{equation}
H
=
H_t
+
H_J
+
H_{\rm SOC}
+
H_2
+
H_{\chi},
\label{eq:Htotal}
\end{equation}
where the individual terms respectively describe nearest-neighbor hopping, noncollinear exchange coupling, intrinsic spin-orbit coupling (SOC), next-nearest-neighbor hopping, and an emergent chiral flux associated with orbital Berry phases.

\subsection{Real-Space Hamiltonian}

The nearest-neighbor (NN) hopping term is
\begin{equation}
H_t
=
-t
\sum_{\langle ij\rangle,\alpha}
c_{i\alpha}^\dagger
c_{j\alpha}
+
{\rm H.c.},
\label{eq:Ht}
\end{equation}
where
$c_{i\alpha}^\dagger$
creates an electron with spin
$\alpha=\uparrow,\downarrow$
on site $i$.
This term generates the characteristic kagome electronic structure consisting of Dirac cones and a nearly dispersionless flat band.

Noncollinear altermagnetism is introduced through the exchange coupling
\begin{equation}
H_J
=
J
\sum_i
c_i^\dagger
(\mathbf{M}_i\cdot\boldsymbol{\sigma})
c_i,
\label{eq:HJ}
\end{equation}
where
$\boldsymbol{\sigma}=(\sigma_x,\sigma_y,\sigma_z)$
denotes the Pauli matrices.
We consider the coplanar $120^\circ$ magnetic texture for the three sites  in the unit cell and atomic magnetization as:
\begin{align}
\mathbf{M}_A
&=
M(1,0,0),
\\
\mathbf{M}_B
&=
M\left(
-\frac12,
\frac{\sqrt3}{2},
0
\right),
\\
\mathbf{M}_C
&=
M\left(
-\frac12,
-\frac{\sqrt3}{2},
0
\right),
\end{align}
which satisfies
\begin{equation}
\mathbf{M}_A
+
\mathbf{M}_B
+
\mathbf{M}_C
=
0.
\end{equation}
The magnetic order is therefore fully compensated and carries no net magnetization despite explicitly breaking time-reversal symmetry $\mathcal T$.

Intrinsic SOC is modeled by a Kane--Mele-type term~\cite{kane2005quantum}:
\begin{equation}
H_{\rm SOC}
=
i\lambda
\sum_{\langle\!\langle ij\rangle\!\rangle}
\nu_{ij}\,
c_i^\dagger
\sigma_z
c_j
+
{\rm H.c.},
\label{eq:HSOC}
\end{equation}
where
$\nu_{ij}=\pm1$
depends on the chirality of the second-neighbor hopping path.
This term preserves lattice translations while generating topological gaps near Dirac crossings.

To introduce particle--hole asymmetry and a dispersion of the
flat-band-derived manifold, we include a same-sublattice
longer-range hopping,
\begin{equation}
H_2
=
-t_2
\sum_{\mathbf R,\mu,\alpha}
\sum_{n=1}^{3}
\left(
c_{\mathbf R\mu\alpha}^{\dagger}
c_{\mathbf R+\mathbf a_n,\mu\alpha}
+\mathrm{H.c.}
\right),
\end{equation}
where $\mu=A,B,C$. This particular hopping is diagonal in the
sublattice space and produces a common momentum-dependent energy
shift of all bands.

Finally, we introduce a nearest-neighbor imaginary bond order,
\begin{equation}
H_\chi
=
i\chi
\sum_{\langle ij\rangle,\alpha}
\xi_{ij}^{\chi}\,
c_{i\alpha}^{\dagger}c_{j\alpha}
+\mathrm{H.c.},
\label{eq:Hchi}
\end{equation}
where $\xi_{ji}^{\chi}=-\xi_{ij}^{\chi}=\pm1$ specifies the
orientation of the loop-current pattern shown in Fig.~1(a).
The total hopping on an oriented bond is therefore
\begin{equation}
t_{ij}^{\rm eff}
=
-t+i\chi\xi_{ij}^{\chi}.
\end{equation}
The gauge-invariant orbital flux through a plaquette $p$ is
\begin{equation}
\Phi_p
=
\arg
\prod_{\langle ij\rangle\in p}
t_{ij}^{\rm eff}.
\end{equation}
Ordinary time reversal maps $\chi\rightarrow-\chi$, so the two
signs of $\chi$ describe orbital-current domains with opposite
circulation.

The parameter $\chi$ should not be interpreted as an externally
imposed uniform magnetic flux. Rather, it represents the amplitude
of a time-reversal-odd loop-current, or imaginary bond-order,
parameter. A possible microscopic origin can be understood by
considering a nonlocal density-density interaction,
\begin{equation}
H_{\mathrm{int}}
=
\sum_{ij} V_{ij} n_i n_j .
\end{equation}
Introducing the bond expectation value
\begin{equation}
Q_{ij}
=
\sum_{\sigma}
\left\langle
c_{i\sigma}^{\dagger}c_{j\sigma}
\right\rangle ,
\end{equation}
a Fock-channel mean-field decoupling generates an effective
bond-dependent hopping. For a complex bond order of the form
\begin{equation}
Q_{ij}
=
Q_{ij}^{\mathrm{R}}
+
i\nu_{ij}Q_{ij}^{\mathrm{I}},
\end{equation}
its imaginary component produces
\begin{equation}
H_{\mathrm{LC}}
=
i\sum_{ij,\sigma}
\chi_{ij}\nu_{ij}
c_{i\sigma}^{\dagger}c_{j\sigma}
+\mathrm{H.c.},
\qquad
\chi_{ij}\propto V_{ij}Q_{ij}^{\mathrm{I}} ,
\end{equation}
which is the loop-current contribution retained phenomenologically
in Eq.~(10). The corresponding order parameter is proportional to
$\operatorname{Im}\langle c_i^\dagger c_j\rangle$ and describes
circulating equilibrium bond currents. It is odd under time reversal,
and the two signs $\chi$ and $-\chi$ label orbital-current domains
with opposite circulation.

Extended-Hubbard studies of kagome systems have shown that nonlocal
Coulomb interactions, particularly near van Hove filling, can
stabilize imaginary bond order and loop-current phases
\cite{dong2023loop}. Experimental observations of time-reversal-symmetry breaking in kagome metals such as AV$_3$Sb$_5$ \cite{mielke2022time} have also motivated loop-current descriptions. In the present
work, we do not calculate the self-consistent formation of this
order. Instead, $\chi$ is treated as a phenomenological symmetry-breaking
field in order to isolate its consequences for the Berry curvature
of the coplanar compensated magnetic state.

\subsection{Momentum-space representation}

We choose the three basis positions as
\begin{equation}
\mathbf r_A=0,\qquad
\mathbf r_B=\frac{\mathbf a_1}{2},\qquad
\mathbf r_C=\frac{\mathbf a_2}{2},
\end{equation}
and use the Fourier convention
\begin{equation}
c_{\mathbf R\mu\alpha}
=
\frac{1}{\sqrt N}
\sum_{\mathbf k}
e^{i\mathbf k\cdot(\mathbf R+\mathbf r_\mu)}
c_{\mathbf k\mu\alpha}.
\end{equation}
In the basis
\begin{equation}
\Psi_{\mathbf k}
=
(
c_{\mathbf kA\uparrow},
c_{\mathbf kA\downarrow},
c_{\mathbf kB\uparrow},
c_{\mathbf kB\downarrow},
c_{\mathbf kC\uparrow},
c_{\mathbf kC\downarrow}
)^T,
\end{equation}
the Bloch Hamiltonian is
\begin{equation}
\mathcal H(\mathbf k)
=
\mathcal H_t(\mathbf k)
+
\mathcal H_2(\mathbf k)
+
\mathcal H_J
+
\mathcal H_{\rm SOC}(\mathbf k)
+
\mathcal H_\chi(\mathbf k).
\end{equation}

Defining
\begin{equation}
k_n=\mathbf k\cdot\mathbf a_n,
\qquad
\mathbf a_3=\mathbf a_1-\mathbf a_2,
\end{equation}
the nearest-neighbor hopping becomes
\begin{equation}
\mathcal H_t(\mathbf k)
=
-2t
\sum_{n=1}^{3}
\cos\frac{k_n}{2}\,
\Lambda_n\otimes\mathbb I_2,
\end{equation}
where
\begin{align}
\Lambda_1&=
\begin{pmatrix}
0&1&0\\
1&0&0\\
0&0&0
\end{pmatrix},
&
\Lambda_2&=
\begin{pmatrix}
0&0&1\\
0&0&0\\
1&0&0
\end{pmatrix},
&
\Lambda_3&=
\begin{pmatrix}
0&0&0\\
0&0&1\\
0&1&0
\end{pmatrix}.
\end{align}

For the translation-preserving loop-current pattern shown in
Fig.~1(a), the directed NN hoppings satisfy
$\xi_{AB}^{\chi}=+1$, $\xi_{AC}^{\chi}=-1$, and
$\xi_{BC}^{\chi}=+1$. The corresponding Bloch contribution is
\begin{equation}
\mathcal H_\chi(\mathbf k)
=
2\chi
\sum_{n=1}^{3}
\cos\frac{k_n}{2}\,
\widetilde\Lambda_n\otimes\mathbb I_2,
\end{equation}
where
\begin{align}
\widetilde\Lambda_1&=
\begin{pmatrix}
0&i&0\\
-i&0&0\\
0&0&0
\end{pmatrix},
&
\widetilde\Lambda_2&=
\begin{pmatrix}
0&0&-i\\
0&0&0\\
i&0&0
\end{pmatrix},
\\
\widetilde\Lambda_3&=
\begin{pmatrix}
0&0&0\\
0&0&i\\
0&-i&0
\end{pmatrix}.
\end{align}

The second-neighbor SOC bonds connect the sublattice pairs
$AB$, $AC$, and $BC$ through the displacement vectors
\begin{align}
\mathbf d_1^{(2)}
&=
\mathbf a_2-\frac{\mathbf a_1}{2},
\\
\mathbf d_2^{(2)}
&=
\mathbf a_1-\frac{\mathbf a_2}{2},
\\
\mathbf d_3^{(2)}
&=
\frac{\mathbf a_1+\mathbf a_2}{2}.
\end{align}
For the chirality convention
$\nu_{AB}^{\rm SO}=+1$,
$\nu_{AC}^{\rm SO}=-1$, and
$\nu_{BC}^{\rm SO}=+1$, the SOC term becomes
\begin{align}
\mathcal H_{\rm SOC}(\mathbf k)
=
2\lambda
\Bigg[
&
\cos\left(k_2-\frac{k_1}{2}\right)
\widetilde\Lambda_1
+
\cos\left(k_1-\frac{k_2}{2}\right)
\widetilde\Lambda_2
\nonumber\\
&
+
\cos\left(\frac{k_1+k_2}{2}\right)
\widetilde\Lambda_3
\Bigg]\otimes\sigma_z.
\end{align}
The same-sublattice longer-range hopping produces
\begin{equation}
\mathcal H_2(\mathbf k)
=
-2t_2
\left[
\cos k_1+\cos k_2+\cos k_3
\right]\mathbb I_6.
\end{equation}
Finally, introducing $\Delta=JM$, the exchange contribution is
\begin{equation}
\mathcal H_J
=
\Delta
\begin{pmatrix}
\sigma_x&0&0\\
0&
-\frac12\sigma_x+\frac{\sqrt3}{2}\sigma_y
&0\\
0&0&
-\frac12\sigma_x-\frac{\sqrt3}{2}\sigma_y
\end{pmatrix}.
\end{equation}

\begin{figure*}[t]
    \centering
    \includegraphics[width=0.99\linewidth]{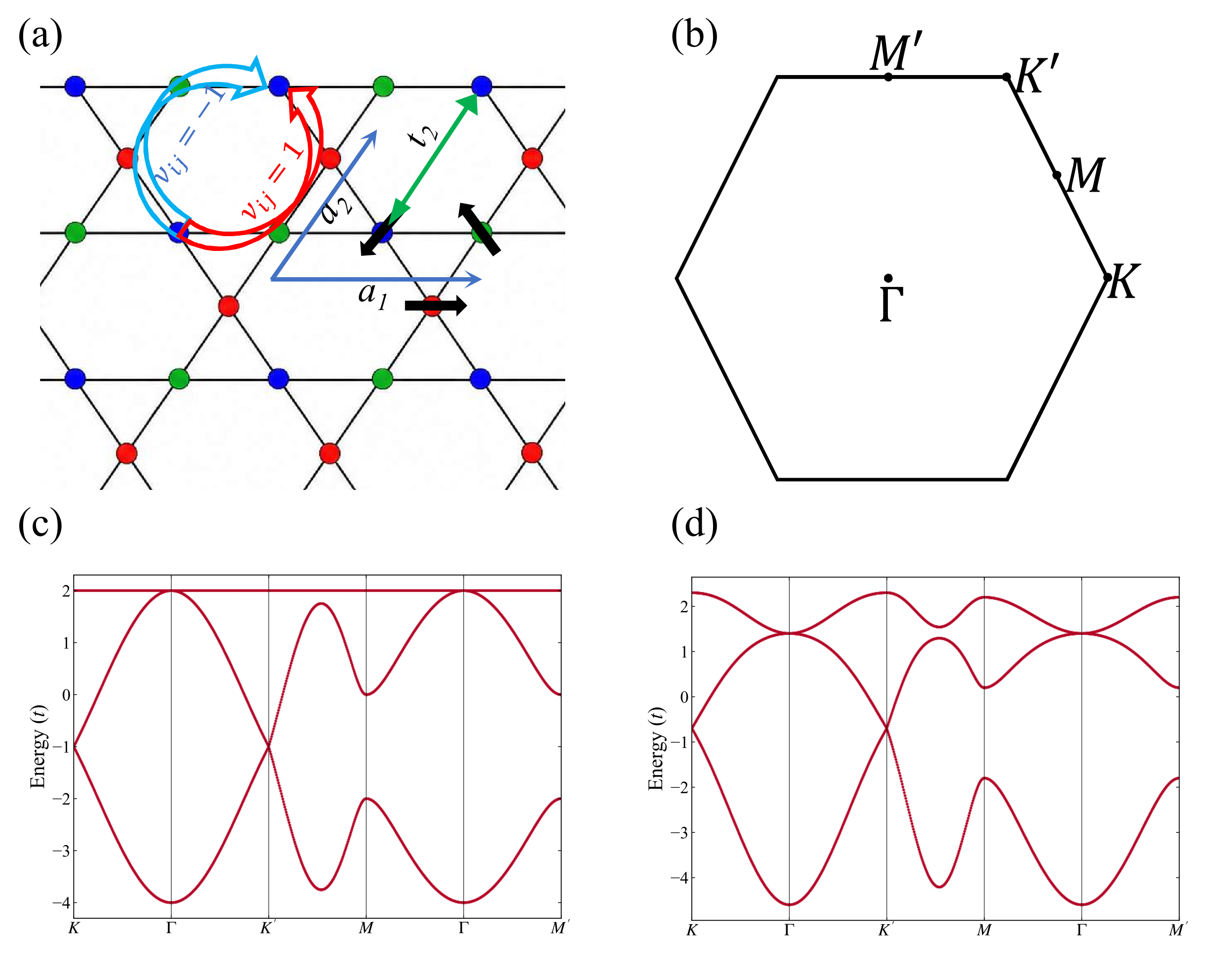}
    \caption{
(a) Schematic kagome lattice illustrating the
three atoms per unit cell, colored with red (A site), green (B site) and blue (C site), the Kane-Mele term, the NNN hopping and the spin arrangement entering the model Hamiltonian. (b)
First Brillouin zone of the kagome lattice with the high-symmetry points indicated. (c) Electronic band structure of the ideal kagome model with only nearest‑neighbor hopping, showing the two dispersive bands and the characteristic flat band at $E=2t$.
(d) Band structure including a finite second‑neighbor hopping $t_2=0.1$, which weakly disperses the flat band while preserving all lattice symmetries. 
}
\label{Figure1}
\end{figure*}

\subsection{Symmetry classification of the noncollinear magnetic state}

Because the extension of altermagnetism to noncollinear magnetic
structures requires an explicit symmetry analysis, we first classify
the coplanar $120^\circ$ state employed in the present model. The
three local moments can be written in the general form
\begin{equation}
\mathbf{M}_{\ell}
=
M
\left[
\cos\left(\phi+\frac{2\pi\ell}{3}\right),
\sin\left(\phi+\frac{2\pi\ell}{3}\right),
0
\right],
\qquad
\ell=0,1,2 ,
\end{equation}
where $\phi$ denotes the global in-plane phase of the magnetic
texture. They satisfy
\begin{equation}
\sum_{\ell=0}^{2}\mathbf{M}_{\ell}=0,
\end{equation}
and hence the spin magnetization is fully compensated.

For the kagome basis, 
spatial inversion maps each sublattice onto itself modulo a lattice
translation.
Because spin is an axial vector, inversion does not reverse the local
magnetic moments. The magnetic texture therefore preserves
$\mathcal P$. In contrast, time reversal transforms
\begin{equation}
\mathcal T:\mathbf M_{\ell}\rightarrow-\mathbf M_{\ell}.
\end{equation}
No lattice translation restores the original $q=0$ magnetic
configuration, and consequently both $\mathcal T$ and
$\mathcal{PT}$ are broken:
\begin{equation}
\mathcal P\in\mathcal G_M,\qquad
\mathcal T\notin\mathcal G_M,\qquad
\mathcal{PT}\notin\mathcal G_M.
\end{equation}

The magnetic structure is therefore fully compensated, inversion
symmetric, time-reversal breaking, and $\mathcal{PT}$ breaking.
Within the symmetry classification of noncollinear altermagnets,
these properties identify the state as an S-type altermagnet~\cite{cheong2025altermagnetism}. For
the high-symmetry orientation of the $120^\circ$ order, the
corresponding magnetic point group is
$6^{\prime}/m^{\prime}m^{\prime}m$. This is the standard
three-sublattice kagome representative of a strong S-type
noncollinear altermagnet.

In the nonrelativistic limit, the magnetic structure is more
appropriately characterized by joint spin-space and real-space
operations. In particular, the system is invariant under the
simultaneous rotation
\begin{equation}
\mathcal C_3
=
\left[
C_{3z}^{\mathrm{spin}}
\parallel
C_{3z}^{\mathrm{space}}
\right],
\end{equation}
which cyclically transforms
\begin{equation}
A\rightarrow B\rightarrow C\rightarrow A,
\qquad
\mathbf M_A\rightarrow\mathbf M_B
\rightarrow\mathbf M_C\rightarrow\mathbf M_A .
\end{equation}
In the sublattice--spin basis, this operation is represented by
\begin{equation}
\mathcal U_3
=
P_{ABC}\otimes
\exp\left(-i\frac{\pi}{3}\sigma_z\right),
\end{equation}
and obeys
\begin{equation}
\mathcal U_3 H(\mathbf k)\mathcal U_3^\dagger
=
H(C_{3z}\mathbf k).
\end{equation}
It therefore relates spin-polarized states at symmetry-related
momenta, rather than producing opposite-spin partners at the same
generic momentum.

The coplanar Hamiltonian also possesses the antiunitary operation
\begin{equation}
\mathcal A
=
\mathcal P U_z(\pi)\mathcal T,
\qquad
U_z(\pi)=\exp\left(-i\frac{\pi}{2}\sigma_z\right).
\end{equation}
Time reversal reverses the in-plane moments, while $U_z(\pi)$
rotates them back to their original directions. Inversion returns
$-\mathbf k$ to $\mathbf k$, such that
\begin{equation}
\mathcal A H(\mathbf k)\mathcal A^{-1}
=
H(\mathbf k).
\end{equation}
Importantly,
\begin{equation}
\mathcal A^2=+1,
\end{equation}
so this operation does not generate a Kramers degeneracy. The
bands can therefore be nondegenerate and spin polarized at a
generic momentum despite the presence of this residual
antiunitary symmetry.

The exchange-induced nondegeneracy and momentum-dependent spin
polarization occur already for $\lambda=0$. Accordingly, the
magnetic state is a strong, rather than weak, S-type noncollinear
altermagnet. Throughout this work, the term ``altermagnet'' is
used in this generalized noncollinear sense.

\begin{table*}[t]
\caption{Symmetry classification of the coplanar $120^\circ$
kagome magnetic state.}
\begin{ruledtabular}
\begin{tabular}{lcc}
Property or operation & Status & Consequence\\
\hline
$\sum_{\ell}\mathbf M_{\ell}$ & $0$
& Full spin compensation\\
$\mathcal P$ & Preserved
& $E_n(\mathbf k)=E_n(-\mathbf k)$\\
$\mathcal T$ & Broken
& Magnetic time-reversal breaking\\
$\mathcal{PT}$ & Broken
& Same-$\mathbf k$ spin degeneracy not enforced\\
$[C_{3z}^{\rm spin}\Vert C_{3z}^{\rm space}]$
& Preserved
& Relates spin textures in rotated momentum sectors\\
$\mathcal A=\mathcal P U_z(\pi)\mathcal T$
& Preserved, $\mathcal A^2=+1$
& No Kramers degeneracy; Berry-curvature constraint\\
SOC-free spin splitting & Finite
& Strong noncollinear altermagnet\\
\end{tabular}
\end{ruledtabular}
\end{table*}

\section{Results}

Before introducing the altermagnetic exchange field, it is instructive to establish the baseline electronic structure of the kagome lattice in the absence of all spin-dependent terms. By setting $J=\lambda=\chi=0$, the Hamiltonian reduces to a spin-degenerate nearest-neighbor tight-binding model. Diagonalizing the $3\times3$ sublattice block yields two dispersive bands,
\begin{equation}
E_{\pm}(\mathbf{k}) = t\!\left[-1 \pm \sqrt{4\cos^{2}\!\frac{k_{1}}{2} + 4\cos^{2}\!\frac{k_{2}}{2} + 4\cos^{2}\!\frac{k_{3}}{2}}\right],
\end{equation}
together with a perfectly flat band at $E_{\mathrm{flat}} = 2t$. As expected for a non-magnetic system preserving both time-reversal ($\mathcal{T}$) and spatial inversion ($\mathcal{I}$) symmetries, all bands remain exactly spin-degenerate, and the total magnetization vanishes identically.

We consider the first Brillouin zone of the system, which is plotted in Fig.~1(b).
The resulting spectrum, shown in Fig.~1(c), exhibits characteristic Dirac touchings at the $K$ and $K'$ points and is fully symmetric under the point group of the hexagonal Brillouin zone, including the mappings $K\!\leftrightarrow\!K'$ and $M\!\leftrightarrow\!M'$. The exactly flat band reflects the destructive interference of electron wavefunctions on the frustrated kagome geometry, leading to a quenched kinetic energy and infinite effective mass in the ideal limit.

To illustrate the effect of realistic band reconstruction, we include a real second-neighbor hopping term $t_{2}$, which contributes a momentum-dependent diagonal shift
\begin{equation}
\varepsilon_{2}(\mathbf{k}) = -2t_{2}\!\left[\cos k_{1} + \cos k_{2} + \cos k_{3}\right].
\end{equation}
For $t_{2}=0.1$, as depicted in Fig.~1(d), this term breaks the destructive interference condition, thereby introducing a finite dispersion to the previously flat band. Furthermore, the $t_2$ term breaks particle-hole symmetry, resulting in an energy shift of the Dirac points relative to the band minima at $\Gamma$. Despite these structural modifications, the system maintains its non-magnetic character; spin degeneracy is preserved across the entire Brillouin zone, as indicated by the vanishing $\langle S_x \rangle$ polarization (grey color) along all high-symmetry paths. This baseline confirms that any subsequent spin-splitting or momentum-dependent polarization can be attributed solely to the altermagnetic exchange and topological terms.

\begin{figure*}[t]
    \centering
    \includegraphics[width=0.99\linewidth]{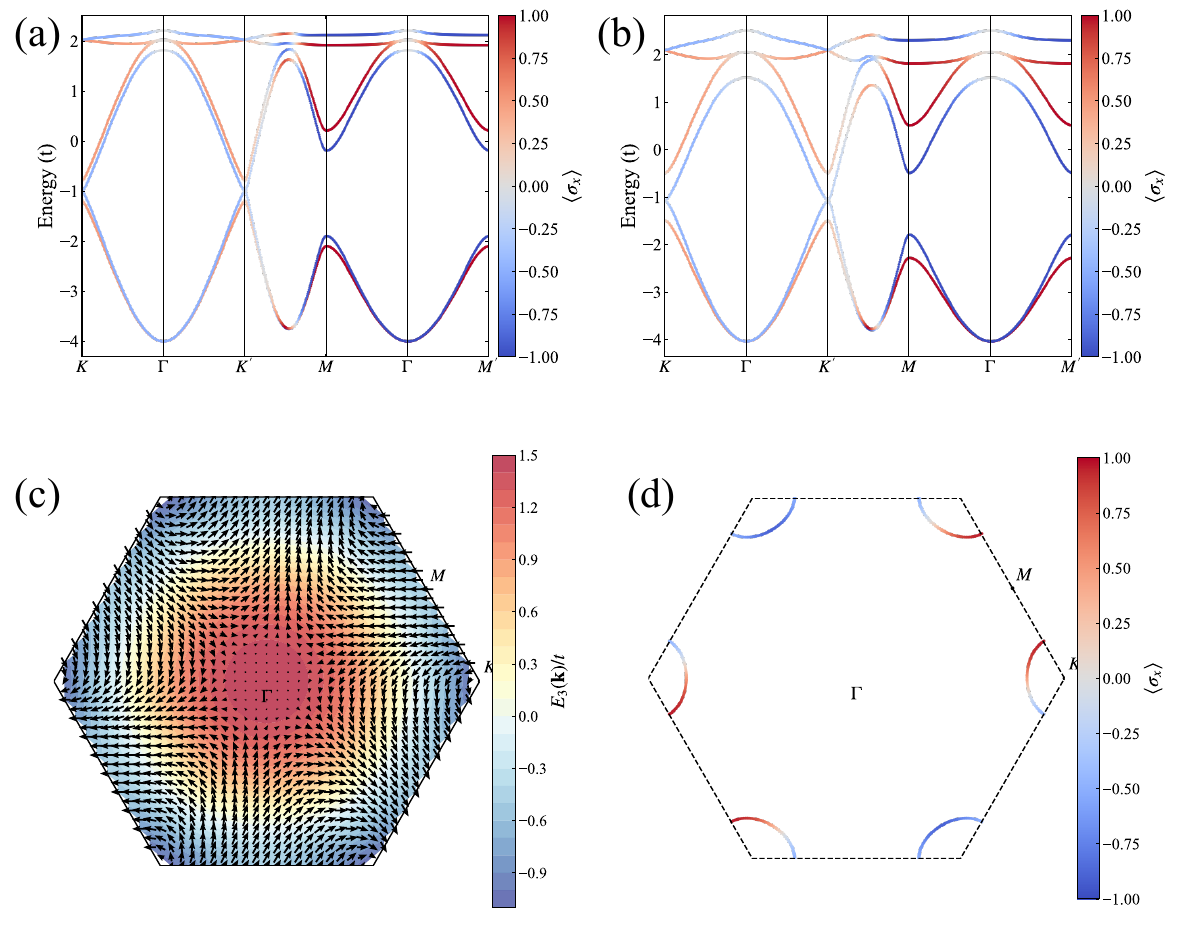}
    \caption{Spin-resolved band structures of the kagome lattice in the presence of a 
coplanar $120^{\circ}$ altermagnetic exchange field for (a) $J=0.2$ and (b) $J=0.5$, 
respectively, with $t_{2}=\lambda=\chi=0$ projected on $\langle S_{x}\rangle$. The exchange coupling lifts the 
spin degeneracy across most of the Brillouin zone and produces a pronounced, 
momentum-dependent spin-splitting characteristic of altermagnetism. 
(c) Momentum-space spin texture $\langle \boldsymbol{S}(\mathbf{k})\rangle$ 
of the third band for $J=0.5$, showing large in-plane polarization whose 
orientation alternates between symmetry-related momenta. 
(d) Corresponding Fermi surface colored by $\langle S_{x}\rangle$, 
illustrating the compensated yet strongly anisotropic spin structure that 
defines the altermagnetic phase.
}
\label{Figure2}
\end{figure*}

To isolate the intrinsic effect of the noncollinear exchange field, we first consider the minimal model with
\(
t_{2}=\lambda=\chi=0
\),
such that the electronic structure is governed solely by nearest-neighbor hopping and the coplanar $120^\circ$ exchange texture. Figures~2(a) and 2(b) show the evolution of the kagome band structure for exchange couplings
\(
J=0.2
\)
and
\(
J=0.5
\),
respectively.

In the absence of exchange coupling, the kagome lattice exhibits the well-known Dirac bands together with a perfectly flat band at
\(
E=2t
\).
Introducing the noncollinear exchange field fundamentally reconstructs this spectrum. Even without relativistic spin--orbit coupling, the magnetic texture lifts the spin degeneracy over large portions of the Brillouin zone, producing strongly momentum-dependent spin splitting. The spin polarization of the bands, represented by the color scale corresponding to
\(
\langle S_x\rangle
\),
reveals the emergence of oppositely polarized electronic branches despite the absence of any net magnetization.

This behavior originates from the symmetry properties of the coplanar $120^\circ$ magnetic state. The noncollinear order breaks time-reversal symmetry while preserving lattice translations and threefold rotational symmetry, thereby removing the antiunitary symmetry constraints that would otherwise enforce Kramers-like spin degeneracy. As a consequence, the exchange coupling alone generates anisotropic momentum-space spin splitting characteristic of altermagnetic systems.

A notable feature appears near the $\Gamma$ point, where the Dirac-derived bands remain grouped into nearly degenerate pairs, while the flat-band manifold undergoes substantial exchange splitting. Increasing the exchange strength from
\(
J=0.2
\)
to
\(
J=0.5
\)
significantly amplifies the reconstruction of the spectrum. The spin splitting becomes larger throughout the Brillouin zone, and the originally dispersionless flat band evolves into two strongly spin-polarized dispersive branches shifted toward higher and lower energies. Simultaneously, the spin polarization approaches nearly saturated values
\(
\langle S_x\rangle \approx \pm1
\),
indicating that the exchange field dominates over the kinetic energy scale.

Importantly, the spin splitting is highly anisotropic in momentum space. Symmetry-related regions such as $M$ and $M'$ exhibit opposite spin polarization, while the magnitude of the splitting varies strongly along different momentum directions. This anisotropic spin-momentum locking distinguishes the present noncollinear phase from conventional ferromagnetic exchange splitting, where the spin separation is approximately momentum independent.

To further characterize the altermagnetic phase, Fig.~2(c) presents the momentum-resolved spin texture
\(
\langle \mathbf S\rangle_{\mathbf k}
\)
for the third band at
\(
J=0.5
\).
The spin texture displays a pronounced noncollinear winding throughout the Brillouin zone, demonstrating that the electronic spin polarization is intrinsically locked to crystal momentum. While the total spin polarization integrated over the Brillouin zone vanishes identically due to magnetic compensation, the local spin expectation values remain finite and rotate continuously around the high-symmetry points. In particular, the spin polarization is strongly suppressed near the $\Gamma$ point but becomes enhanced near the zone boundaries, where the exchange-induced band reconstruction is maximal. The resulting momentum-space spin pattern provides direct evidence for nonrelativistic altermagnetic spin splitting generated purely by the noncollinear exchange field.

The influence of this spin splitting on the itinerant electronic states is further illustrated in Fig.~2(d), which shows the Fermi surface projected onto the
\(
\langle S_x\rangle
\)
component for
\(
J=0.5
\).
The Fermi surface consists of multiple spin-polarized pockets distributed near the edges of the hexagonal Brillouin zone. Unlike conventional antiferromagnets, where opposite-spin states remain degenerate at each momentum point, the present system exhibits strongly spin-polarized Fermi contours with alternating positive and negative spin character. The local spin polarization changes sign between symmetry-related momentum sectors, reflecting the underlying $120^\circ$ magnetic point-group symmetry.

This spin-polarized yet magnetically compensated Fermi surface is a defining hallmark of altermagnetism. The coexistence of vanishing net magnetization and strong momentum-dependent spin polarization suggests that kagome noncollinear altermagnets may provide an efficient platform for spin-current generation, anisotropic magnetotransport, and unconventional Hall responses even in the absence of relativistic spin--orbit coupling.

The inclusion of intrinsic SOC introduces the
relativistic effect to the compensated kagome altermagnetic state, as
illustrated in Fig.~3.  In this section, we consider the regime
$\chi=0$, such that the electronic structure is governed solely by the
interplay between the coplanar $120^\circ$ exchange texture and intrinsic
SOC.  Although SOC substantially reconstructs the band crossings and modifies the momentum-space spin texture, the system remains topologically compensated
due to a hidden antiunitary symmetry that survives in the coplanar magnetic
phase.

Figure~3(a) presents the spin-resolved band structure for
$\lambda=0.1$.  Compared with the purely nonrelativistic spectrum shown in
Fig.~2(b), SOC clearly reconstructs the electronic bands near the Dirac
crossings at the $K$ and $K'$ points and introduces additional level
repulsion close to the $\Gamma$ point.  The originally nearly degenerate
Dirac-derived manifolds become separated by SOC-induced hybridization gaps,
while the flat-band sector acquires a weak momentum-dependent deformation.
Despite these relativistic modifications, the electronic states remain strongly
spin split throughout large regions of the Brillouin zone.  The color
projection of $\langle S_x\rangle$ demonstrates that the momentum-dependent
altermagnetic spin polarization survives even in the presence of sizeable SOC,
indicating that the dominant spin splitting still originates from the
noncollinear exchange texture rather than from relativistic effects.


To characterize the momentum-space topology generated by the combined action
of noncollinear altermagnetism and SOC, we evaluate the Berry curvature within
the Kubo linear-response formalism.  Since the present system is metallic, the physically relevant Berry curvature is obtained by summing over all occupied
states below the Fermi energy,
\begin{equation}
\Omega_z(\mathbf k)
=
\sum_n
f\big(E_n(\mathbf k)-E_F\big)\,
\Omega_z^n(\mathbf k),
\end{equation}
where
$f(E)$
is the Fermi--Dirac distribution and
$E_F$
denotes the Fermi level.

The band-resolved Berry curvature is computed using
\begin{equation}
\Omega_z^n(\mathbf{k})
=
-2\,\mathrm{Im}
\sum_{m\neq n}
\frac{
\langle n|v_x|m\rangle
\langle m|v_y|n\rangle
}{
(E_n-E_m)^2+\eta^2
},
\label{eq:berrykubo}
\end{equation}
where
$v_\mu=\partial H/\partial k_\mu$
is the velocity operator and
$\eta=0.03$ 
is a small numerical broadening introduced to regularize near-degenerate
states.  The velocity operators are evaluated numerically using finite
momentum derivatives of the Bloch Hamiltonian.

A remarkable and highly nontrivial feature of the coplanar kagome
altermagnetic phase is such that the Berry curvature remains identically zero
throughout the Brillouin zone, even in the simultaneous presence of
noncollinear magnetic order and intrinsic SOC, as shown in Fig.~3(b).
At first sight, this result appears counterintuitive because the exchange field
explicitly breaks ordinary time-reversal symmetry and produces pronounced
momentum-dependent spin splitting.  Nevertheless, the vanishing Berry curvature originates from a hidden antiunitary symmetry that survives in the
strictly coplanar magnetic state.

The essential observation is that all magnetic moments $\mathbf M_i$, where $i$ is the index of the atoms, remain confined to the kagome plane,
\begin{equation}
\mathbf M_i=(M_i^x,M_i^y,0),
\end{equation}
with no out-of-plane spin component.  Under the time-reversal operator,
\begin{equation}
\mathcal T:\quad
\mathbf M_i
\rightarrow
-\mathbf M_i,
\end{equation}
the magnetic texture changes sign.  However, the transformed configuration can
be exactly restored by a global $\pi$ spin rotation around the $z$ axis,
\begin{equation}
C_{2z}:\quad
(M_x,M_y,M_z)
\rightarrow
(-M_x,-M_y,M_z).
\end{equation}
Consequently, the combined antiunitary operation
\begin{equation}
\mathcal S=\mathcal T C_{2z}
\end{equation}
leaves the full Hamiltonian invariant even in the presence of intrinsic SOC.
The coplanar altermagnetic state therefore retains an effective antiunitary
symmetry despite explicitly breaking conventional time-reversal symmetry.

This hidden symmetry imposes a stringent constraint on the Berry curvature.
Under time reversal,
\begin{equation}
\Omega_z(\mathbf k)
\rightarrow
-\Omega_z(-\mathbf k),
\end{equation}
whereas the twofold spin rotation $C_{2z}$ transforms momentum according to
\begin{equation}
(k_x,k_y)
\rightarrow
(-k_x,-k_y),
\end{equation}
while leaving the out-of-plane Berry curvature component unchanged,
\begin{equation}
\Omega_z(\mathbf k)
\rightarrow
\Omega_z(-\mathbf k).
\end{equation}
Combining both operations yields
\begin{equation}
\mathcal T C_{2z}:\quad
\Omega_z(\mathbf k)
\rightarrow
-\Omega_z(\mathbf k).
\end{equation}
Since the Hamiltonian is invariant under $\mathcal T C_{2z}$, the Berry
curvature must satisfy
\begin{equation}
\Omega_z(\mathbf k)
=
-\Omega_z(\mathbf k),
\end{equation}
which immediately enforces
\begin{equation}
\Omega_z(\mathbf k)=0
\end{equation}
for all crystal momenta.

From a geometric perspective, this result is directly connected to the
vanishing scalar spin chirality of the coplanar $120^\circ$ texture,
\begin{equation}
\kappa_{ijk}
=
\mathbf S_i
\cdot
(
\mathbf S_j
\times
\mathbf S_k
).
\end{equation}
Because all spins lie within the same plane,
\begin{equation}
\kappa_{ijk}=0,
\end{equation}
no spin-chirality-induced emergent magnetic field is generated.
The coplanar kagome altermagnet therefore realizes an unusual electronic phase in which strong
momentum-dependent spin splitting coexists with a completely vanishing Berry
curvature and anomalous Hall response.

This behavior sharply distinguishes the present system from conventional
noncollinear kagome antiferromagnets~\cite{nayak2016large},
where spin canting or noncoplanarity breaks the hidden
$\mathcal T C_{2z}$ symmetry and immediately generates large Berry-curvature
hot spots and anomalous Hall conductivity.  In contrast, the ideal coplanar
altermagnetic state studied here remains topologically compensated even in the
presence of sizeable SOC.  The system, therefore, constitutes a symmetry-protected altermagnetic metal: it exhibits giant nonrelativistic spin
splitting while remaining completely Berry-curvature silent.

\begin{figure*}[t]
    \centering
    \includegraphics[width=0.99\linewidth]{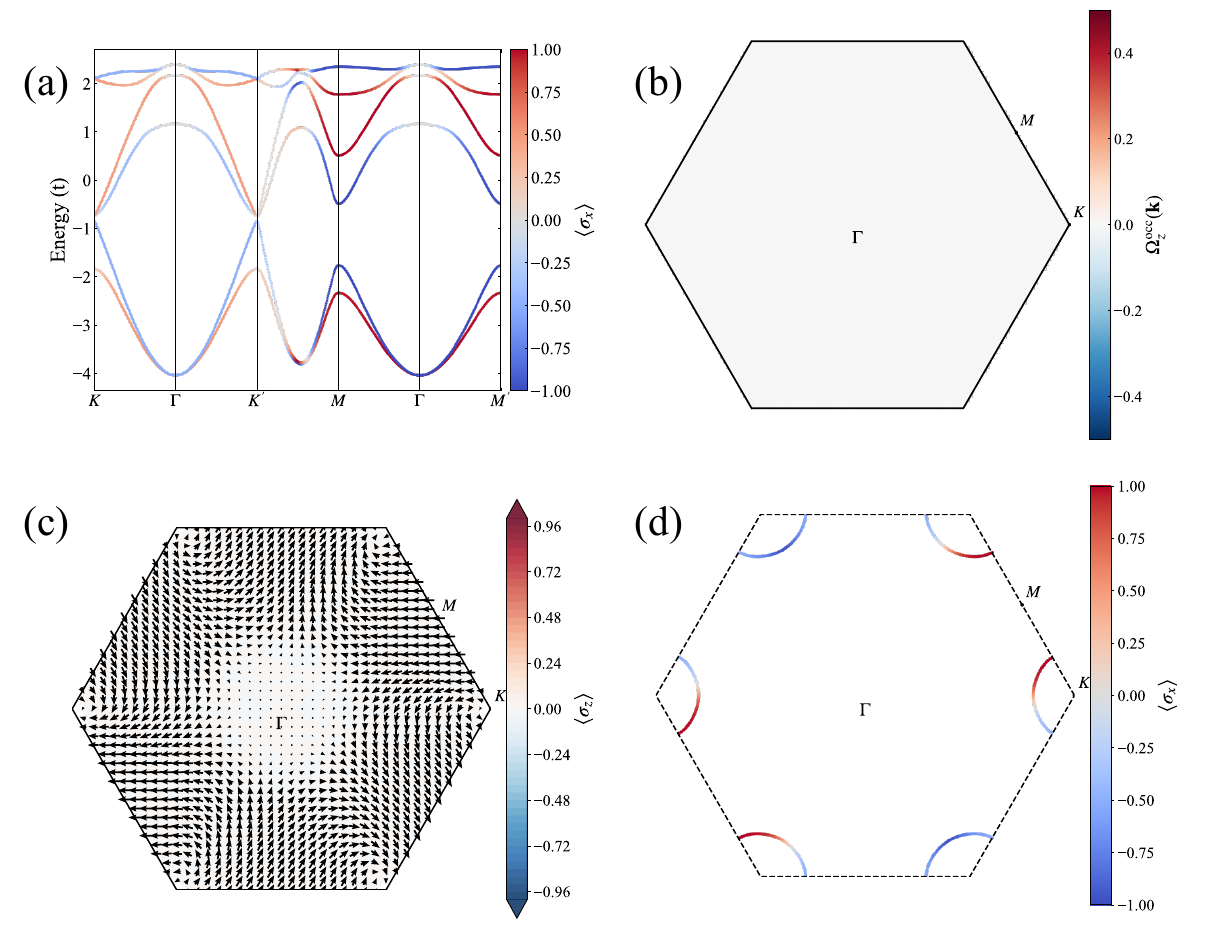}
    \caption{(a) Spin-resolved band structures of the kagome lattice in the presence of a 
spin-orbit, $\lambda=0.1$,  and with $J=0.5$, 
 projected on $\langle S_{x}\rangle$. 
(b) Berry curvature of the model, demonstrating that the SOC cannot produce finite Berry curvature in the present model. (c) Spin texture in the presence of the SOC. The background shows the $\langle S_{z}\rangle$, indicating the SOC cannot align spin along the z-direction.
(d) Corresponding Fermi surface colored by $\langle S_{x}\rangle$ in the presence of SOC. Other parameters are $t_2=\chi=0$.
}
\label{Figure3}
\end{figure*}

The emergence of finite Berry curvature consequently requires an additional
mechanism capable of breaking the hidden $\mathcal T C_{2z}$ symmetry.  In the
present model, this role is naturally played by the chiral flux term
$H_\chi$, which mimics interaction-driven orbital circulation and generates an
effective momentum-space magnetic flux.  Once $\chi\neq0$, the antiunitary
compensation is destroyed, allowing Berry-curvature hot spots and anomalous
Hall transport to emerge.  This establishes a clear hierarchy of electronic
phases in the kagome altermagnet: a symmetry-protected coplanar
altermagnetic metal for $\chi=0$, and a topological altermagnetic metal once
orbital chiral currents are activated.

Figure~3(c) further demonstrates that intrinsic SOC does not induce any
appreciable out-of-plane spin polarization in the coplanar altermagnetic
phase.  Although SOC modifies the in-plane spin texture and slightly reshapes
the momentum dependence of
$\langle S_x(\mathbf{k})\rangle$
and
$\langle S_y(\mathbf{k})\rangle$,
the out-of-plane component remains strictly zero within numerical precision.
This behavior again follows directly from the hidden antiunitary symmetry
$\mathcal S=\mathcal T C_{2z}$.  Since the operator $S_z$ is odd under
$\mathcal S$ while the Hamiltonian remains invariant, the expectation value
must satisfy
\begin{equation}
\langle S_z(\mathbf{k})\rangle
=
-
\langle S_z(\mathbf{k})\rangle,
\end{equation}
thereby enforcing
\begin{equation}
\langle S_z(\mathbf{k})\rangle=0
\end{equation}
throughout the Brillouin zone.  Thus, the intrinsic SOC does not produce a canting of the spins out of the kagome plane as long as the chiral flux term remains absent.

Finally, Fig.~3(d) compares the Fermi surface in the presence of SOC with the
SOC-free case shown in Fig.~2(d).  The altermagnetic exchange field already
produces strongly spin-polarized Fermi pockets, while SOC slightly distorts
their geometry and enhances the momentum dependence of the in-plane spin
texture.  Importantly, however, the spin polarization remains strictly
coplanar and no out-of-plane spin canting develops.  The survival of the
hidden antiunitary symmetry $\mathcal T C_{2z}$ therefore protects the system
against both finite Berry curvature and SOC-induced out-of-plane spin
polarization, leaving the kagome altermagnet in a symmetry-protected
topologically compensated metallic state.


We now turn to the regime
\begin{equation}
J\neq0,
\qquad
\chi\neq0,
\qquad
\lambda=0,
\qquad
t_2=0,
\end{equation}
where the electronic structure is governed solely by nearest-neighbor hopping,
the coplanar $120^\circ$ altermagnetic exchange texture, and the emergent
chiral flux term $H_\chi$.  In this limit, the Hamiltonian reduces to
\begin{equation}
H
=
H_t+H_J+H_\chi.
\end{equation}
Unlike the SOC-driven case discussed in Fig.~3, the present phase is entirely
nonrelativistic: all topological effects originate purely from orbital gauge
structure rather than spin--orbit entanglement. It should be emphasized that the orbital-flux term $H_\chi$ is independent of the scalar spin chirality of the magnetic texture. Throughout this work, the local moments remain strictly coplanar and therefore satisfy $\kappa_{ijk}=0$. The role of $H_\chi$ is instead to introduce an orbital gauge structure that explicitly breaks the hidden antiunitary symmetry $\mathcal T C_{2z}$ responsible for the Berry-curvature silence of the coplanar altermagnetic phase.

A central question is whether finite Berry curvature can emerge in the absence
of both relativistic SOC and noncoplanar spin chirality.  At first sight, one
might expect the answer to be negative because the magnetic texture remains
strictly coplanar,
\begin{equation}
\kappa_{ijk}
=
\mathbf S_i
\cdot
(
\mathbf S_j
\times
\mathbf S_k
)
=
0,
\end{equation}
and therefore no conventional emergent magnetic field associated with scalar
spin chirality is generated.  Remarkably, however, we find that the chiral
orbital flux term alone is sufficient to produce pronounced Berry curvature
hot spots throughout momentum space, even in the complete absence of SOC.
Figure~4 demonstrates that the resulting phase constitutes a genuine
nonrelativistic topological altermagnetic metal driven entirely by orbital
circulation effects.


The origin of the finite Berry curvature can be understood from symmetry.
For $\chi=0$, the coplanar kagome altermagnet preserves the hidden
antiunitary symmetry
\begin{equation}
\mathcal S
=
\mathcal T C_{2z},
\end{equation}
which was shown in Fig.~3 to enforce
\begin{equation}
\Omega_z(\mathbf k)=0
\end{equation}
identically throughout the Brillouin zone.  The chiral flux term explicitly
breaks this protection.

Under time reversal,
\begin{equation}
\mathcal T H_\chi \mathcal T^{-1}
=
-H_\chi,
\end{equation}
because the hopping amplitudes carry purely imaginary phases.  Meanwhile, the
twofold spin rotation $C_{2z}$ reverses the in-plane magnetic moments but does
not remove the oriented bond chirality encoded in the kagome flux factors
$\nu_{ij}$.  Consequently,
\begin{equation}
\mathcal S H_\chi \mathcal S^{-1}
\neq
H_\chi,
\end{equation}
and the antiunitary compensation responsible for the Berry-curvature silence
of Fig.~3 is destroyed.  Once $\chi\neq0$, no remaining symmetry enforces
\begin{equation}
\Omega_z(\mathbf k)
=
-
\Omega_z(\mathbf k),
\end{equation}
allowing finite local Berry curvature to emerge.


The physical meaning of $H_\chi$ becomes particularly transparent when viewed
as an emergent orbital gauge field. Electrons moving around an elementary
kagome triangle accumulate a geometric phase

\begin{equation}
\Phi_\triangle
=
\arg\!\left(
t_{AB}\, t_{BC}\, t_{CA}
\right),
\end{equation}

where the directed complex hopping amplitudes are

\begin{equation}
t_{AB} = t + i\chi\,\nu_{AB},~  
t_{BC} = t + i\chi\,\nu_{BC}, ~
t_{CA} = t + i\chi\,\nu_{CA}.
\end{equation}
To leading order in $\chi/t$, the accumulated phase is
$\Phi_\triangle \approx \frac{\chi}{t}$
demonstrating that the chiral term generates an effective staggered orbital
flux through the kagome plaquettes. In this sense, $H_\chi$ plays a role
analogous to a Haldane-type gauge field on the kagome lattice, producing local
orbital circulation without requiring either relativistic SOC or noncoplanar
spin textures.

\begin{figure*}[t]
    \centering
    \includegraphics[width=0.99\linewidth]{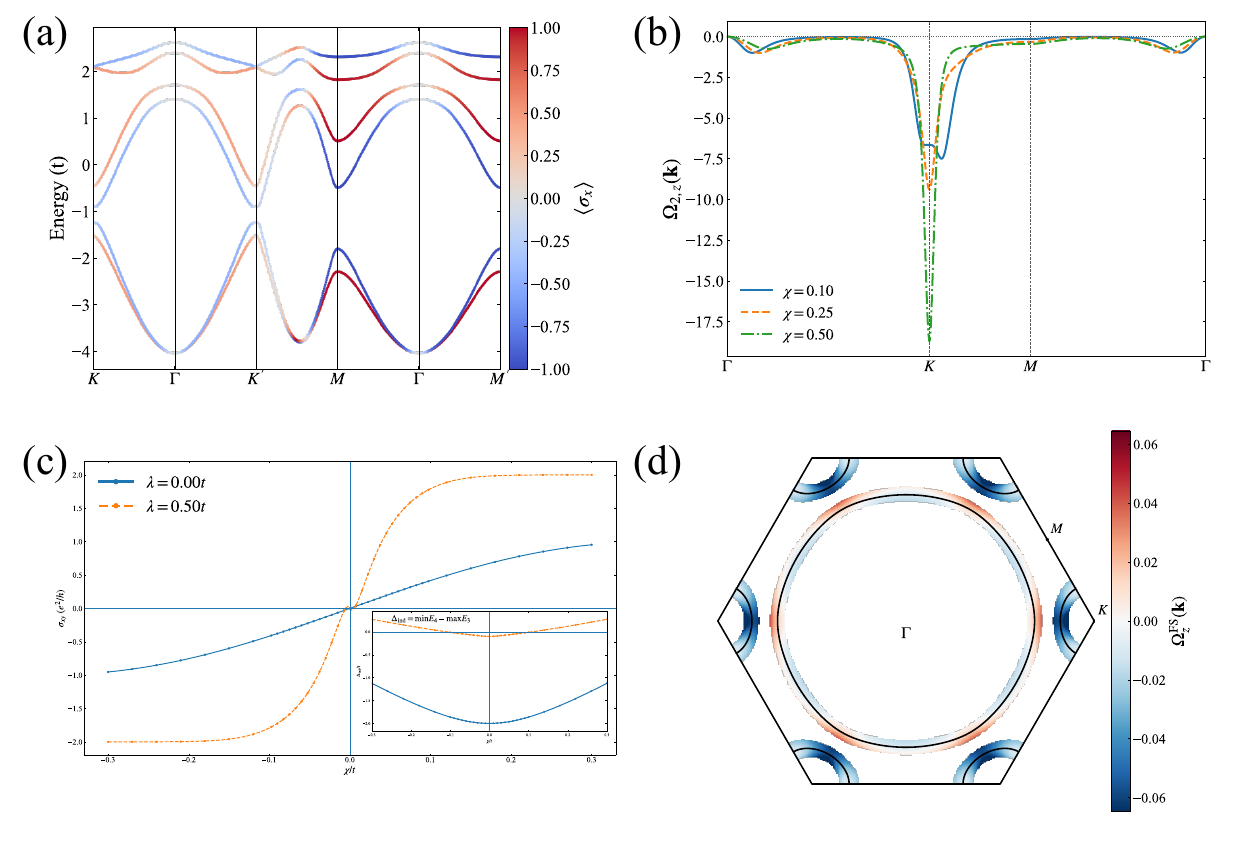}
    \caption{
Electronic topology induced by the chiral orbital flux term $H_\chi$ in the absence of intrinsic spin--orbit coupling ($\lambda=0$).
(a) Spin-resolved band structure for the coplanar kagome altermagnet with finite chiral flux $\chi=0.1$.  The color scale represents the spin polarization
$\langle S_x\rangle$.
(b) Berry curvature
$\Omega_z(\mathbf{k})$
evaluated along the high-symmetry path
$\Gamma$--$K$--$M$--$\Gamma$
for several values of the chiral flux amplitude $\chi$.
Finite Berry-curvature hot spots emerge immediately once $\chi\neq0$ despite the absence of relativistic SOC, demonstrating that orbital gauge flux alone can generate momentum-space topology.
(c) Intrinsic anomalous Hall conductivity
$\sigma_{xy}$
as a function of the chiral flux strength for different SOC amplitudes in a constant filling factor of $n=3$. Inset shows the variation of indirect band gap between band 3 and 4 as a function of $\chi$ for two values of $\lambda$.
(d) Fermi surface projected onto the Berry curvature distribution for
$\chi=0.1$
and
$\lambda=0$ for Band 3 and filling factor of $n=3$.
}
\label{Figure4}
\end{figure*}


The emergence of Berry curvature follows directly from the momentum-space
Hamiltonian
\begin{equation}
\mathcal H(\mathbf k)
=
\mathcal H_t(\mathbf k)
+
\mathcal H_J
+
\mathcal H_\chi(\mathbf k).
\end{equation}
The chiral contribution contains antisymmetric hopping matrices
$\widetilde{\Lambda}_n$ identical in structure to those appearing in the
Kane--Mele SOC term, but acting entirely in the orbital sector rather than
through spin-dependent coupling.  Crucially, evaluating the commutator, we obtain:
\begin{equation}
[
\mathcal H_J,
\mathcal H_\chi(\mathbf k)
]
\neq0,
\end{equation}
so the exchange field and orbital flux cannot be simultaneously diagonalized.
The Bloch eigenstates, therefore, acquire a momentum-dependent complex
sublattice coherence even in the absence of relativistic effects.

To leading order in $\chi$, the Berry curvature of each band responds
linearly to the orbital flux,
\begin{equation}
\Omega_z^n(\mathbf k) \propto \chi,
\end{equation}
with matrix elements controlled by the interference between the velocity
operators and the orbital-flux term $H_\chi$.  The exchange field lifts the
relevant band degeneracies, allowing the orbital flux to generate finite
momentum-space Berry curvature even in the absence of relativistic SOC.


The electronic reconstruction induced by the chiral flux is shown in Fig.~4(a). Compared with the SOC-driven spectrum
of Fig.~3(a), the loop-current term produces a qualitatively
different band evolution. While intrinsic SOC primarily modifies
relativistic hybridization gaps near band crossings, the imaginary
nearest-neighbor hopping associated with $\chi$ reconstructs the
kagome dispersion through complex inter-sublattice hopping
processes. In particular, it modifies the flat-band-derived
manifold and shifts the avoided crossings near the $K$ valleys and
the Brillouin-zone boundaries.

The bands remain strongly spin split because of the underlying
noncollinear altermagnetic exchange field. The color projection of
$\langle \sigma_x\rangle$ demonstrates that the momentum-dependent
spin polarization survives throughout the Brillouin zone even for
$\lambda=0$. The spin splitting is therefore fundamentally
nonrelativistic, whereas the additional band reconstruction arises
from the orbital-current order.


The most direct consequence of the orbital-current order is the
appearance of finite Berry curvature. In the absence of $\chi$, the
antiunitary symmetry of the coplanar magnetic state suppresses the
Berry curvature. The loop-current term breaks this protection and
allows
$\Omega_{n,z}(\mathbf k)\neq0$.
Figure~4(b) displays the Berry curvature of the second band along
the high-symmetry path
$\Gamma$--$K$--$M$--$\Gamma$
for several values of $\chi$. Pronounced Berry-curvature hot spots
develop near the $K$ valley, where the loop-current hopping produces
strong interband mixing and small energy separations. Weaker
features are also visible near other portions of the path.

The magnitude and width of the dominant $K$-valley feature evolve
strongly with increasing $\chi$, reflecting the reconstruction of
the relevant avoided crossings. Importantly, these results are
obtained for $\lambda=0$, demonstrating that relativistic SOC is
not required for the generation of Berry curvature in the present
coplanar kagome altermagnet. Instead, the geometric response is
activated directly by the time-reversal-odd imaginary bond order.


\begin{figure*}[t]
    \centering
    \includegraphics[width=0.99\linewidth]{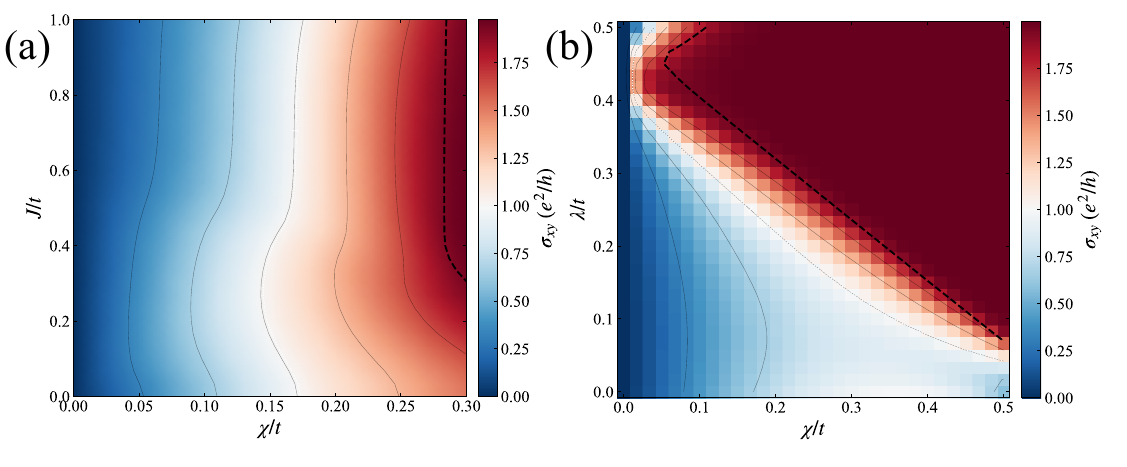}
    \caption{Anomalous Hall conductivity $\sigma_{xy}$ at fixed filling $n_e=3$.
(a) $\sigma_{xy}$ in the $(J,\chi)$ plane for $\lambda=0.25t$.
The Hall response vanishes along $\chi=0$ and increases once the
loop-current order breaks the protecting antiunitary symmetry.
(b) $\sigma_{xy}$ in the $(\lambda,\chi)$ plane for $J=0.5t$.
SOC enhances the loop-current-induced Hall response and drives
$\sigma_{xy}$ toward $2e^2/h$ in the gapped regime. The black dashed
line marks $\Delta_{\mathrm{ind}}=0$, separating metallic and
globally gapped regions.
}
\label{Figure5}
\end{figure*}


To quantify the momentum-integrated response, we evaluate the
intrinsic anomalous Hall conductivity,
\begin{equation}
\sigma_{xy}(\mu)
=
-\frac{e^2}{\hbar}
\sum_n
\int_{\mathrm{BZ}}
\frac{d^2k}{(2\pi)^2}
f\!\left[E_n(\mathbf k)-\mu\right]
\Omega_{n,z}(\mathbf k).
\label{eq:sigma_xy}
\end{equation}
The velocity operators are evaluated analytically as
\begin{equation}
v_\alpha(\mathbf k)
=
\frac{\partial \mathcal H(\mathbf k)}
{\partial k_\alpha}.
\end{equation}
The Brillouin-zone integration is performed over a uniform
reciprocal primitive-cell mesh. For the results shown here, we use
$N_k=151$, a Kubo regularization $\eta=0.01t$, and
$k_{\mathrm B}T=0.02t$. The conductivity is reported in units of
$e^2/h$.

Because varying $\chi$ reconstructs the band energies and therefore
changes the carrier density at fixed chemical potential, the
comparison in Fig.~4(c) is performed at fixed electron filling,
\begin{equation}
n_e
=
\frac{1}{N_k^2}
\sum_{\mathbf k,n}
f\!\left[E_n(\mathbf k)-\mu\right]
=
3 .
\label{eq:filling_fixed}
\end{equation}
The chemical potential is consequently determined independently for
each value of $\chi$ and $\lambda$.

Figure~4(c) shows that reversing the loop-current domain reverses
the Hall response,
\begin{equation}
\sigma_{xy}(-\chi)
=
-\sigma_{xy}(\chi).
\end{equation}
This odd dependence is a direct consequence of the fact that
$\chi$ and $-\chi$ are related by time reversal and correspond to
opposite orbital-current circulations.

For $\lambda=0$, the anomalous Hall conductivity is already finite
and increases continuously with $|\chi|$, reaching values of order
$e^2/h$ in the investigated interval. The response remains
nonquantized because the system stays metallic, as confirmed by the
negative indirect gap shown in the inset. Thus, the loop-current
order alone is sufficient to convert the Hall-silent coplanar
altermagnet into an anomalous Hall metal.

The effect of SOC is qualitatively different from merely activating
the Hall response. For $\lambda=0.5t$, SOC strongly reconstructs the
avoided crossings and enhances the Berry-curvature contribution of
the occupied bands. As $|\chi|$ increases, the indirect gap
\begin{equation}
\Delta_{\mathrm{ind}}
=
\min_{\mathbf k} E_4(\mathbf k)
-
\max_{\mathbf k} E_3(\mathbf k)
\end{equation}
changes sign and becomes positive, as shown in the inset of
Fig.~4(c). The opening of this global gap coincides with the
saturation of the Hall conductivity near
\begin{equation}
\left|\sigma_{xy}\right|
\simeq
2\frac{e^2}{h}.
\end{equation}
This plateau is consistent with a Chern-insulating regime having a
total occupied-band Chern number of magnitude $|C|=2$. In the
small-$|\chi|$ region, before the global gap opens, the system
remains metallic and the Hall conductivity is more sensitive to the
detailed Fermi-surface reconstruction.

These results demonstrate that SOC is not required for a finite
anomalous Hall effect. Rather, $\chi$ removes the
symmetry protection and generates the Hall response, while SOC can
reorganize the Berry-curvature hot spots, open a global topological
gap, and drive the system from an anomalous Hall metal into a
quantized Chern-insulating phase.


Figure~4(d) further illustrates the momentum-space origin of the
Hall response by displaying the Berry curvature in the vicinity of
the Fermi surface. The black contours indicate the Fermi-surface
pockets, while the color scale represents the Fermi-surface-weighted
Berry curvature,
\begin{equation}
\Omega_z^{\mathrm{FS}}(\mathbf k)
=
\frac{
\displaystyle
\sum_n
w_n(\mathbf k)\,
\Omega_{n,z}(\mathbf k)
}{
\displaystyle
\sum_n
w_n(\mathbf k)
},
\end{equation}
where
\begin{equation}
w_n(\mathbf k)
=
\exp\left[
-\frac{
\left(E_n(\mathbf k)-\mu\right)^2
}{
2\delta_E^2
}
\right]
\end{equation}
selects states within a narrow energy window around the chemical
potential.

The Berry curvature is concentrated predominantly along the
spin-split Fermi contours, with especially pronounced contributions
near the Brillouin-zone boundaries and the $K$-centered pockets.
The alternating positive and negative regions reveal that the Hall
response results from an incomplete cancellation of strongly
momentum-dependent Berry-curvature contributions. The finite
integrated conductivity therefore originates not from a uniform
Berry curvature, but from the imbalance among several
Fermi-surface hot spots generated by the loop-current-induced band
reconstruction.

Unlike conventional noncollinear kagome antiferromagnets, where
Berry curvature is often associated with relativistic SOC, weak
spin canting, or a finite scalar spin chirality, the present model
provides a distinct mechanism. In the $\lambda=0$ limit, finite
Berry curvature and anomalous Hall conductivity emerge from orbital
loop-current order within a strictly coplanar magnetic background.
SOC is not required for this activation mechanism, although it can
subsequently open a topological gap and stabilize a quantized Hall
state.

Overall, Fig.~4 establishes a hierarchy of orbital and relativistic
topological responses in the kagome altermagnet. The coplanar
altermagnetic state is initially Hall silent because of its
antiunitary symmetry. Orbital-current order breaks this protection
and produces finite Berry curvature and a nonquantized anomalous
Hall response already in the nonrelativistic limit. SOC then
reconstructs the occupied-band topology and, together with
sufficiently strong loop-current order, drives the system into a
gapped phase with a Hall conductivity approaching
$2e^2/h$. The relativistic and orbital mechanisms can therefore be
distinguished while remaining cooperatively tunable within the same
kagome platform.

Figure~5 summarizes the evolution of the anomalous Hall response as a
function of the exchange coupling, intrinsic spin--orbit interaction, and
loop-current order. Unlike the momentum-resolved Berry-curvature
distributions shown in Figs.~3 and 4, these phase diagrams reveal how the
integrated Hall conductivity evolves across the parameter space of the
model. The calculations are performed at fixed filling $n_e=3$, while the
black dashed curves indicate the closing of the indirect gap,
$\Delta_{\mathrm{ind}}
=
\min_{\mathbf k}E_4(\mathbf k)
-
\max_{\mathbf k}E_3(\mathbf k)
=
0$.

Figure~5(a) presents the anomalous Hall conductivity $\sigma_{xy}$ in the
$(J,\chi)$ parameter space for fixed intrinsic spin--orbit coupling
$\lambda=0.25t$. The conductivity vanishes along $\chi=0$ despite the
presence of finite exchange coupling and SOC. This behavior is consistent
with the symmetry analysis of Fig.~3, where the hidden antiunitary symmetry
$\mathcal{T}C_{2z}$ suppresses the Berry curvature of the coplanar
altermagnetic state. Once the loop-current order is activated, this
protection is removed and a finite anomalous Hall response develops.

For fixed $J$, the conductivity generally increases with $\chi$ and reaches
values close to $2e^2/h$ at sufficiently strong loop-current order. Its
dependence on the exchange coupling is nonmonotonic, with the strongest
response occurring in a broad intermediate-$J$ region. This behavior is
consistent with a competition between exchange-induced band splitting and
loop-current-driven interband hybridization. At intermediate exchange
strength, the reconstructed bands retain sufficiently small energy
separations to generate strong Berry-curvature contributions, whereas a
larger exchange field progressively modifies these avoided crossings and
reduces the Hall response from its maximum value. The map therefore
demonstrates that the exchange field controls not only the altermagnetic
spin splitting but also the efficiency with which the loop-current order
produces an integrated Hall response.

Figure~5(b) shows the Hall conductivity in the $(\lambda,\chi)$ plane for
fixed exchange coupling $J=0.5t$. The conductivity again vanishes along
$\chi=0$, even for sizeable SOC, confirming that intrinsic SOC alone does
not activate the anomalous Hall response in the symmetry-protected
coplanar state. For finite $\chi$, however, increasing SOC substantially
enlarges the region of high Hall conductivity. The response approaches
$2e^2/h$ over a broad part of the parameter space where both $\lambda$ and
$\chi$ are sufficiently large.

The dashed curve in Fig.~5(b) separates the metallic regime from the region
with a positive indirect gap. The development of the globally gapped phase
is accompanied by the saturation of $\sigma_{xy}$ near $2e^2/h$, consistent
with the Chern-insulating regime identified in Fig.~4(c). Below the
gap-opening boundary, the system remains metallic and the Hall response
depends more sensitively on the detailed reconstruction of the Fermi
surface and the Berry-curvature hot spots.

The combined phase diagrams establish a clear hierarchy among the three
energy scales of the model. The exchange coupling $J$ controls the
nonrelativistic altermagnetic spin splitting, the loop-current amplitude
$\chi$ breaks the Hall-protecting antiunitary symmetry and activates Berry
curvature, and the intrinsic SOC $\lambda$ reconstructs the avoided
crossings, enhances the Hall response, and promotes the formation of a
globally gapped topological phase. This separation of magnetic, orbital,
and relativistic contributions is a distinctive feature of the present
kagome altermagnetic model. In particular, finite Berry curvature and
anomalous Hall transport can arise from orbital loop-current order within
a strictly coplanar magnetic background, without requiring either
noncoplanar spin textures or SOC as the primary symmetry-breaking
mechanism.

\begin{figure*}[t]
    \centering
    \includegraphics[width=0.99\linewidth]{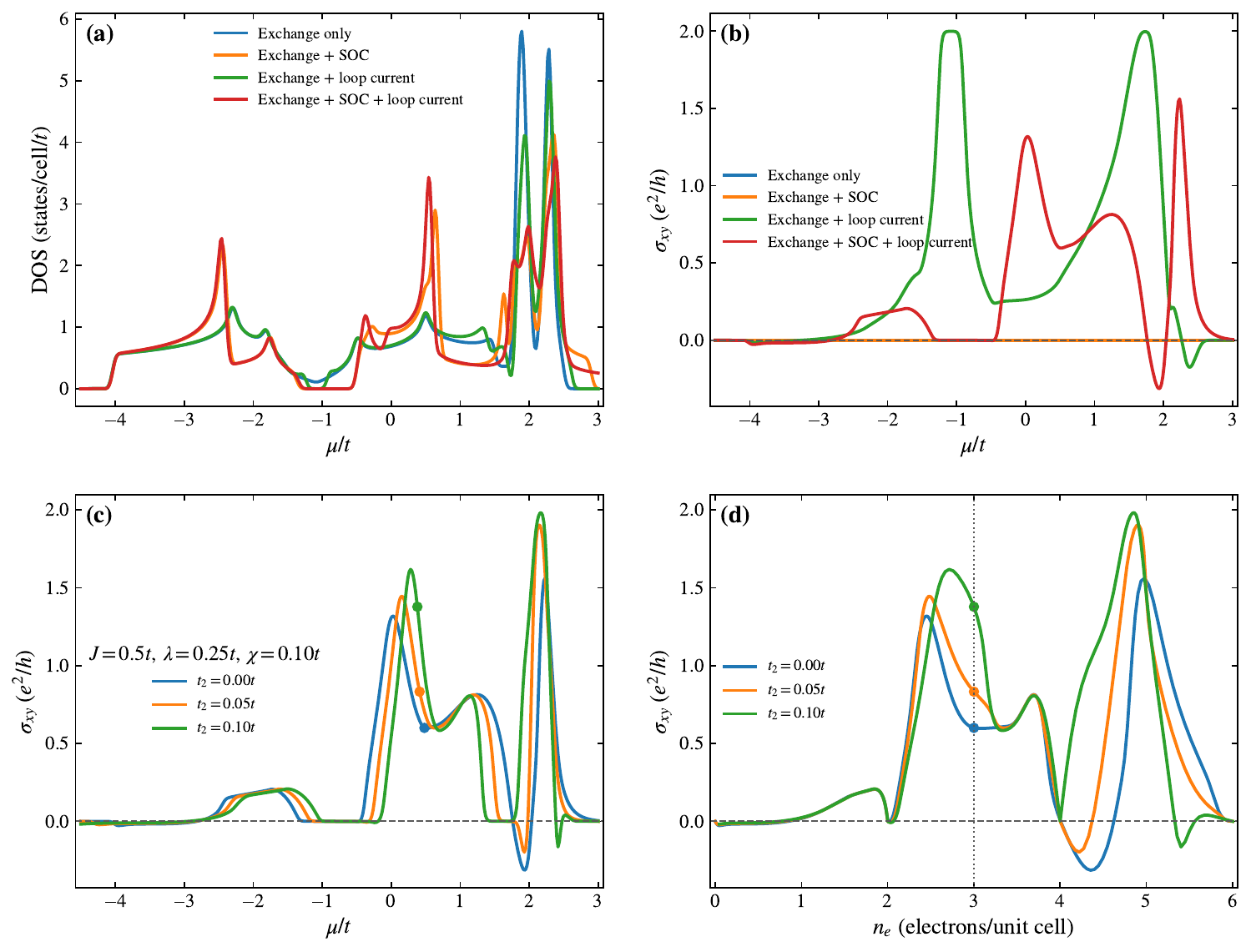}
    \caption{Filling dependence and robustness of the intrinsic anomalous Hall
response in the kagome altermagnetic model.
(a) Density of states as a function of chemical potential for the
exchange-only, exchange-plus-SOC, exchange-plus-loop-current, and
combined exchange--SOC--loop-current regimes.
(b) Corresponding anomalous Hall conductivity
$\sigma_{xy}(\mu)$ in units of $e^2/h$. The phases with
$\chi=0$ remain Hall inactive within numerical precision, whereas
loop-current order generates a finite and strongly
filling-dependent Hall response already in the absence of SOC.
SOC further reconstructs and redistributes the Hall-conductivity
features.
(c) Hall conductivity as a function of chemical potential for
$t_2=0$, $0.05t$, and $0.10t$, with
$J=0.5t$, $\lambda=0.25t$, and $\chi=0.10t$.
(d) The same conductivity represented as a function of the electron
filling $n_e$ per three-site unit cell. Longer-range hopping shifts
the Hall-response features relative to the chemical potential while
preserving the loop-current-induced activation mechanism. The
circles mark the values at fixed filling $n_e=3$.
}
\label{Figure6}
\end{figure*}

Figure~6 examines how the anomalous Hall response evolves with chemical
potential and electron filling and tests its robustness against the
symmetry-preserving longer-range hopping $t_2$. In contrast to the
parameter-space maps of Fig.~5, which were evaluated at fixed filling
$n_e=3$, Fig.~6 resolves the Hall response over the full energy and
filling ranges of the six-band model.

Figure~6(a) shows the density of states for four representative regimes:
exchange only, exchange plus SOC, exchange plus loop-current order, and
the combined exchange--SOC--loop-current state. Both SOC and the
loop-current term reconstruct the kagome spectrum, shifting van Hove
features, modifying the flat-band-derived peaks, and redistributing the
low-density regions associated with avoided crossings and possible
spectral gaps. The combined state exhibits the strongest reconstruction,
reflecting the simultaneous influence of relativistic hybridization and
time-reversal-breaking imaginary bond order.

The corresponding anomalous Hall conductivities are shown in
Fig.~6(b). The exchange-only and exchange-plus-SOC curves remain zero
within numerical precision over the entire chemical-potential range.
This confirms that SOC alone does not remove the antiunitary symmetry
protecting the coplanar altermagnetic state from a finite Hall response.
The vanishing conductivity is therefore not restricted to one particular
filling, but persists throughout the spectrum as long as the
loop-current order is absent.

Once $\chi$ becomes finite, a strongly filling-dependent Hall response
develops. In particular, the exchange-plus-loop-current state already
exhibits large values of $\sigma_{xy}$ without SOC, demonstrating that
the orbital-current order itself is sufficient to activate Berry
curvature and anomalous Hall transport. The conductivity varies strongly
with $\mu$ because different Berry-curvature hot spots enter and leave
the occupied manifold as the Fermi level is swept through the bands.
Broad regions in which $\sigma_{xy}$ approaches $2e^2/h$ coincide with
strongly depleted density of states and indicate parameter intervals
close to a gapped topological regime. These regions should be interpreted
as quantized Chern plateaus only when a positive indirect gap and the
corresponding occupied-band Chern number are verified.

SOC does not simply produce a uniform enhancement of the Hall response.
Instead, comparison between the loop-current-only and combined curves
shows that SOC redistributes the Berry curvature across the spectrum,
shifts the Hall-conductivity peaks and plateaus, and can produce both
enhancement and suppression depending on the filling. This behavior is
consistent with the picture established in Figs.~4 and 5: the
loop-current order breaks the Hall-protecting symmetry and activates the
geometric response, while SOC reconstructs the avoided crossings and
modifies how the activated Berry curvature is distributed among the
bands.

Figures~6(c) and 6(d) investigate the robustness of this response against
the longer-range hopping $t_2$, Eq.31.
Because this term is proportional to the identity matrix in the internal
sublattice--spin space, it does not modify the eigenvectors or the
band-resolved Berry curvature at a fixed momentum. It does, however,
shift the band energies in a momentum-dependent manner and therefore
changes which states are occupied at a given chemical potential or
electron filling.

As shown in Fig.~6(c), increasing $t_2$ shifts the principal Hall
features along the chemical-potential axis and changes their relative
amplitudes. When the same data are represented as a function of the
actual electron filling in Fig.~6(d), the curves retain the same overall
sequence of Hall-active regions, although their precise positions and
magnitudes remain filling dependent. The circles indicate the values at
$n_e=3$, thereby connecting Fig.~6 directly to the fixed-filling results
of Figs.~4 and 5. The conductivity also approaches zero in the empty- and
fully-filled-band limits, as required because no states, or all six
bands, contribute to the occupied manifold.

Overall, Fig.~6 demonstrates that the anomalous Hall effect is not tied
to a fine-tuned chemical potential. The loop-current order produces a
finite response over broad filling intervals, while SOC and longer-range
hopping reshape its detailed energy dependence without restoring the
Hall-silent symmetry. The activation mechanism is therefore robust:
the exchange field establishes the nonrelativistic altermagnetic band
splitting, the loop-current order generates the symmetry breaking
required for finite Berry curvature, and SOC together with additional
band-dispersion terms controls the location and strength of the resulting
Hall-active regions.

\begin{table*}[t]
\caption{
Hierarchy of electronic regimes in the kagome altermagnetic model.
The final column refers to the intrinsic Hall response at generic filling;
quantization requires a global gap at the chosen filling.
}
\begin{ruledtabular}
\begin{tabular}{lcccll}
Regime &
$J$ &
$\lambda$ &
$\chi$ &
Berry curvature &
Hall response
\\
\hline
Nonmagnetic kagome metal
& $0$ & $0$ & $0$
& zero
& zero
\\
Coplanar altermagnet
& $\neq0$ & $0$ & $0$
& zero
& zero
\\
SOC-renormalized altermagnet
& $\neq0$ & $\neq0$ & $0$
& zero
& zero
\\
Loop-current altermagnet
& $\neq0$ & $0$ & $\neq0$
& finite
& finite, generally nonquantized
\\
SOC-assisted loop-current phase
& $\neq0$ & $\neq0$ & $\neq0$
& finite
& strong; quantized when gapped
\end{tabular}
\end{ruledtabular}
\label{tab:phase_hierarchy}
\end{table*}

\section{Discussion}
\label{sec:discussion}

The results establish a minimal framework in which nonrelativistic
altermagnetic spin splitting, Berry-curvature activation, and anomalous
Hall transport are controlled by distinct terms in the Hamiltonian.
The exchange coupling $J$ produces the momentum-dependent spin splitting
of the coplanar $120^\circ$ state, the loop-current amplitude $\chi$
introduces a time-reversal-odd imaginary bond order, and the intrinsic
spin--orbit coupling $\lambda$ reconstructs the resulting avoided
crossings. This separation is summarized in
Table~\ref{tab:phase_hierarchy} and provides a transparent way to
distinguish magnetic, orbital, and relativistic contributions to the
electronic response.

For $\chi=0$, the antiunitary symmetry $\mathcal A$ identified above
remains intact. It forces the Berry curvature, and consequently the
intrinsic anomalous Hall conductivity, to vanish in both the
exchange-only and exchange-plus-SOC regimes. The SOC term can therefore
modify the dispersion and open local hybridization gaps without, by
itself, activating a Hall response. This result emphasizes that
spin-split bands are not sufficient for an anomalous Hall effect: the
magnetic and crystalline symmetries of the Bloch Hamiltonian remain
decisive.

The situation changes qualitatively when the loop-current order becomes
finite. The imaginary nearest-neighbor bond order breaks $\mathcal A$
and generates finite Berry curvature already at $\lambda=0$. In contrast
to the earlier compensated interpretation, the corrected model supports
a sizable, filling-dependent anomalous Hall conductivity in the
nonrelativistic limit. SOC is therefore not the fundamental source of
the Hall response. Instead, it redistributes the Berry curvature,
reconstructs the relevant avoided crossings, and can convert the
loop-current-induced anomalous Hall metal into a globally gapped phase.
At filling $n_e=3$, the Hall conductivity approaches
$2e^2/h$ once the indirect gap opens, consistently with an occupied-band
Chern number of magnitude two.

This mechanism also separates orbital chirality from scalar spin
chirality. The coplanar texture satisfies
\begin{equation}
\kappa_{ijk}
=
\mathbf S_i\cdot
\left(
\mathbf S_j\times\mathbf S_k
\right)
=
0,
\end{equation}
throughout the parameter range considered here. The Hall-active phase
instead originates from directed complex hopping amplitudes and the
associated circulation around lattice plaquettes. It therefore does
not require spin canting or noncoplanar magnetism. Recent symmetry-based
work has likewise stressed the importance of separating
magnetization-driven Hall contributions from zero-magnetization
contributions permitted by crystal and magnetic symmetries
\cite{bu2026disentanglinganomaloushalleffect}. In the present model, this distinction is
particularly direct because $\chi$ activates the Hall response without
creating a net magnetic moment.

The filling dependence in Fig.~6 further shows that the Hall
conductivity is controlled by the occupation of Berry-curvature hot
spots rather than by a single universal energy scale.  It therefore
does not change the eigenvectors or the band-resolved Berry curvature at
fixed momentum, but it shifts the band energies and changes which states
are occupied at a given chemical potential or filling. The resulting
displacement and reshaping of the Hall-active intervals demonstrate that
the activation mechanism is symmetry robust, while the magnitude and
sign of the integrated response remain sensitive to band filling.

The model is relevant to the broader family of kagome magnets, where
large Berry-curvature-driven Hall responses have been observed in
noncollinear antiferromagnets such as Mn$_3$Sn and Mn$_3$Ge and in
ferromagnetic kagome metals such as Fe$_3$Sn$_2$ and
Co$_3$Sn$_2$S$_2$
\cite{nayak2016large,nayak2016large,ye2018massive,
Kiyohara2016PRB}. These materials, however, realize different magnetic
symmetries and microscopic mechanisms, and the present single-orbital
model should not be regarded as a quantitative description of any one
compound. Recent work on cubic Mn$_3$Ge, for example, demonstrates that
intrinsic noncoplanar order and competing magnetic states can generate
additional Hall phenomenology beyond the strictly coplanar mechanism
considered here \cite{hu2026competingmagneticstatesnoncoplanar}.

The minimal character of the model also defines its principal
limitations. Real transition-metal kagome systems contain several
active $d$-orbital manifolds, crystal-field splittings, interorbital
hybridization, interlayer coupling, and orbital-dependent SOC.
Experiments and multiorbital calculations for CoSn demonstrate that
different orbital sectors can host substantially different Dirac gaps
and flat-band dispersions \cite{liu2020orbital}, while multiorbital kagome
models support a richer set of topological bands than the single-orbital
limit \cite{okamoto2022topological}. Correlations provide an additional
renormalization: DFT+DMFT calculations for Mn$_3$Sn find
orbital-dependent mass enhancement, reconstructed Fermi surfaces, and
substantial shifts of topological crossings relative to the Fermi
energy \cite{yu2022correlated}. Such effects can modify the position,
magnitude, and sign of the Hall response and may also provide a
microscopic route to the imaginary bond order represented
phenomenologically by $\chi$.

These material-specific complications do not alter the symmetry
criterion derived here. Any perturbation that preserves $\mathcal A$
cannot activate the intrinsic Hall response of the $\chi=0$ phase.
Once loop-current order breaks $\mathcal A$, Berry curvature becomes
symmetry allowed, but its detailed momentum distribution and integrated
value depend on orbital content, filling, and interactions. A
quantitative treatment of a specific compound will therefore require a
Wannier-based multiorbital Hamiltonian and, where correlations are
important, an interacting approach such as DFT+DMFT.

\begin{acknowledgments}
M.B.T. acknowledges the funding support by  Narodowa Agencja Wymiany Akademickiej (NAWA) under the ULAM program with project number BPN/ULM/2025/1/00156/U/00001.
M.B.T. acknowledges the funding support by Iran National Science Foundation (INSF) under project No.4043973. This research was supported by the Foundation for Polish Science project “MagTop” no. FENG.02.01-IP.05-0028/23 co-financed by the European Union from the funds of Priority 2 of the European Funds for a Smart Economy Program 2021–2027 (FENG). We further acknowledge access to the computing facilities of the Interdisciplinary Center of Modeling at the University of Warsaw, Grant g91-1418, g91-1419, g96-1808, g96-1809 and g103-2540 for the availability of high-performance computing resources and support. We acknowledge the access to the computing facilities of the Poznan Supercomputing and Networking Center, Grants No. pl0807, pl0267-01, pl0365-01 and pl0471-01.
\end{acknowledgments}

\medskip

\appendix

\bibliography{references}
\end{document}